\documentclass[aps,twocolumn,prl,superscriptaddress,amsmath,showpacs,tightenlines,nofootinbib]{revtex4-2}
\usepackage{epsfig,graphicx,times}
\usepackage{amstext}
\usepackage{amsmath}            
\usepackage{amssymb}            
\usepackage{graphicx}           
\usepackage{latexsym}
\usepackage{bm}
\usepackage{color}
\makeatletter
\@ifundefined{textcolor}{}
{\definecolor{BLACK}{gray}{0}
 \definecolor{WHITE}{gray}{1}
 \definecolor{RED}{rgb}{1,0,0}
 \definecolor{GREEN}{rgb}{0,1,0}
 \definecolor{BLUE}{rgb}{0,0,1}
 \definecolor{CYAN}{cmyk}{1,0,0,0}
 \definecolor{MAGENTA}{cmyk}{0,1,0,0}
 \definecolor{YELLOW}{cmyk}{0,0,1,0}
}

\newcommand{\vari}[1]{{#1}}

\newcommand{\Liu}[1]{{#1}}

\newcommand{\MS}[1]{{#1}}
\newcommand{\MSB}[1]{{#1}}

\makeatother
\begin{document}
\title{Optomechanical Anti-lasing with Infinite Group Delay at a Phase Singularity}

\author{Yulong Liu}
\affiliation{Beijing Academy of Quantum Information Sciences, Beijing 100193, China}
\affiliation{Department of Applied Physics, Aalto University, P.O. Box 15100, FI-00076 Aalto, Finland}

\author{Qichun Liu}
\affiliation{Beijing Academy of Quantum Information Sciences, Beijing 100193, China}

\author{Shuaipeng Wang}
\affiliation{Quantum Physics and Quantum Information Division, Beijing Computational Science Research Center, Beijing 100193, China}

\author{Zhen Chen}
\affiliation{Beijing Academy of Quantum Information Sciences, Beijing 100193, China}

\author{Mika A.~Sillanp\"{a}\"{a}}
\email{mika.sillanpaa@aalto.fi}
\affiliation{Department of Applied Physics, Aalto University, P.O. Box 15100, FI-00076 Aalto, Finland}

\author{Tiefu Li}
\email{litf@tsinghua.edu.cn}
\affiliation{School of Integrated Circuits and Frontier Science Center for Quantum Information, Tsinghua University, Beijing 100084, China}
\affiliation{Beijing Academy of Quantum Information Sciences, Beijing 100193, China}
\date{\today}
\begin{abstract}
Singularities which symbolize abrupt changes and exhibit extraordinary behavior are of a broad interest. We experimentally study optomechanically induced singularities in a compound system consisting of a three-dimensional aluminum superconducting cavity and a metalized high-coherence silicon nitride membrane resonator. Mechanically-induced coherent perfect absorption and anti-lasing occur simultaneously under a critical optomechanical coupling strength. Meanwhile, the phase around \MSB{the} cavity resonance undergoes an abrupt $\pi$-phase transition, which further flips the phase slope in \MSB{the} frequency dependence. The observed infinite-discontinuity in the phase slope defines a singularity, at which the group velocity is dramatically changed. Around the singularity, an abrupt transition from an infinite group advance to delay is demonstrated by measuring \MSB{a} Gaussian-shaped waveform propagating. Our experiment may broaden the scope of realizing extremely long group \MSB{delays} by taking advantage of singularities.
\end{abstract}
\maketitle

Electromagnetically induced transparency (EIT), where a transmission window is induced within an absorption resonance, arises from the destructive quantum interference of different excitation pathways~\cite{fleischhauer2005electromagnetically,liu2017electromagnetically,gu2017microwave}. As a counterpart of EIT, electromagnetically induced absorption (EIA) accompanied by a substantial enhancement of the absorption rate of a probe field~\cite{akulshin1998electromagnetically,lezama1999electromagnetically,taichenachev1999electromagnetically, lipsich2000absorption,goren2003electromagnetically,sheng2011all} also has received considerable attention due to its versatile applications such as enhancing photodetection~\cite{romero2009microwave,peropadre2011approaching,akhlaghi2015waveguide}, giant Faraday rotation~\cite{floess2017plasmonic}, and high-precise quantum sensing~\cite{thaicharoen2019electromagnetically,liao2020microwave,jadoon2020multiphoton}. In contrast to EIT introduced by destructive quantum interference, the observed EIA arises due to the constructive interference between different excitation pathways~\cite{dimitrijevic2011coherent,zhang2015electromagnetically,limonov2017fano,yelikar2020analogue}. In particular, the incident electromagnetic field can be perfectly absorbed by precisely controlling the constructive interference~\cite{agarwal2015photon,agarwal2016perfect,wang2017interference,wei2018coherent,xiong2020coherent}. Coherent perfect absorption (CPA) is achieved when the dip of EIA spectrum \MSB{approaches} zero~\cite{baranov2017coherent,zhang2017observation,wu2016coherent,wu2017perfect}. Intriguing applications through the CPA include synthetic reflectionless media~\cite{roger2015coherent,li2018tunable,berkhout2019perfect}, \vari{time-reversed lasers}~\cite{chong2010coherent,wan2011time,stone2011gobbling,wong2016lasing} and random anti-lasing~\cite{pichler2019random}.

The above intriguing applications are closely related to the amplitude responses of light transmission. Meanwhile, the phase also exhibits interesting properties, e.g., EIT and EIA with unique behaviors of dispersion~\cite{boyd2009slow,ian2010tunable,li2011experimental,peng2014and,lu2018optomechanically,zhao2021phase} could lead to group delay (slow light) or advance (fast light), respectively. Classical analogs for EIT and EIA have been simultaneously demonstrated in a variety platforms such as plasmonics~\cite{taubert2012classical}, metamaterials~\cite{tassin2012electromagnetically}, and optical resonators~\cite{yang2017realization}. Similar to EIT and EIA realized through mechanical effects of light~\cite{agarwal2010electromagnetically,huang2011electromagnetically,qu2013phonon}, optomechanically induced transparency (OMIT) and absorption (OMIA) are also caused by destructive and constructive pathway interferences~\cite{weis2010optomechanically,safavi2011electromagnetically,karuza2013optomechanically,dong2013transient,dong2015brillouin,kim2015non}, respectively. In the microwave domain, OMIT~\cite{teufel2011circuit,ockeloen2019sideband,massel2012multimode,zhou2013slowing,fan2015cascaded} and OMIA~\cite{hocke2012electromechanically,massel2011microwave,fong2014microwave,zhang2016cavity} have been demonstrated in electromechanical devices where a superconducting microwave resonator couples to a mechanical resonator realized as a capacitor~\cite{aspelmeyer2014cavity,xiong2018fundamentals,liu2018cavity}. Integrating electromechanics with solid-state qubits, e.g., superconducting qubits~\cite{rodrigues2019coupling,pirkkalainen2013hybrid,bera2021large,lecocq2015resolving,zoepfl2020single}, leads to a promising hybrid architecture for a quantum repeater~\cite{mirhosseini2020superconducting}. The parametric optomechanical coupling has a great tunability. \vari{\MS{Change of probe light transmission} from absorption to transparency \MS{or} amplification has been experimentally observed by tuning the power of a blue-detuned sideband pump~\cite{hocke2012electromechanically}. The probe light could be completely absorbed and fast-slow light conversion occurs at a critical pump power~\cite{chen2021tunable}.} The \MSB{amplitude} and \MSB{phase of} the cavity output photons are both measurable. Therefore, \MSB{optomechanical systems provide} an ideal experimental platform to study how CPA affects the phase evolution, and allow us to study fast-slow light transition, and then explore \MSB{an} ultralong group delay around this phase singularity.

\begin{figure}[ptb]
\includegraphics[scale=0.5]{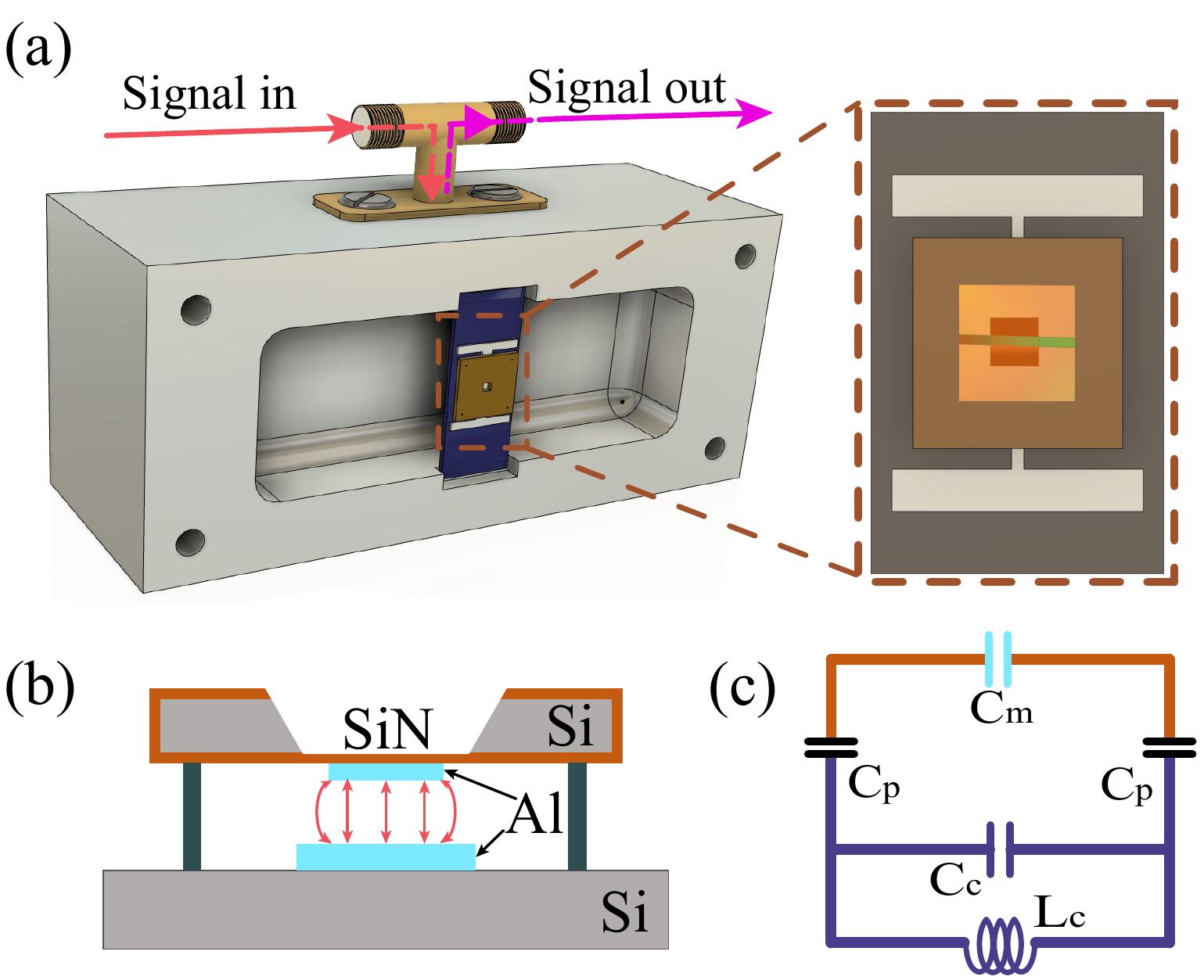}
\caption{(a) The 3D microwave cavity and \MSB{a} mechanically compliant \MSB{capacitor.} Microwave signals couple in and out of the cavity through a common connector. (b) Schematic showing the placement of the metalized membrane over the antenna pads. (c) \Liu{Lumped}-element model consists of cavity capacitance $C_{c}$, inductance $L_{c}$, antenna-cavity coupling capacitance $C_{p}$, and the mechanical capacitance $C_{m}$.}
\label{fig1}
\end{figure}

\textit{The device.}---As shown in Fig.~\ref{fig1}(a), the cavity electromechanical device consists of a three-dimensional (3D) superconducting aluminum (Al) cavity and a mechanically compliant capacitor. The external coupling strength is adjusted by the protrusion length of the SMA connector's pin inside the cavity. The dark-red interference fringes within the narrow capacitor gap \MSB{indicate} that the distance is at a few hundred nanometers level. \MSB{The} schematic in Fig.~\ref{fig1}(b) shows how the metalized silicon-nitride (SiN) membrane is flip-chip mounted over the bottom antenna pads. Detailed fabrication and packaging processes are presented in the Supplementary Material~\cite{Liu2021antilasing}. The H-shaped antenna is used to enhance \MSB{the} electromechanical coupling. The choice of \MSB{a} high-stress SiN membrane as the mechanical resonator is motivated by its high-Q performance at millikelvin temperatures~\cite{yuan2015large,yuan2015silicon,noguchi2016ground,ghadimi2018elastic,tsaturyan2017ultracoherent,maccabe2020nano}.

The equivalent lumped-element model of the cavity and membrane capacitor is shown in Fig.~\ref{fig1}(c). The vibration of the SiN membrane changes the capacitance of the circuit resonator and causes a change in its resonant frequency. Therefore, a dispersive optomechanical coupling is formed. A coherent pump tone with frequency $\Omega_{c}$ and amplitude $\xi$ is used to enhance the optomechanical coupling~\cite{aspelmeyer2014cavity}. In this experiment, the pump frequency is red-detuned \MSB{from} the cavity resonant frequency ($\omega_{c}$) \vari{by the} mechanical frequency ($\omega_{m}$), i.e., $\Omega_{c}=\omega_{c}-\omega_{m}$. A weak coherent probe field with frequency $\Omega_{p}$ and amplitude $\varepsilon$ is used to measure the transmission response. Working in the rotating frame at the pump-tone frequency, the linearized Hamiltonian is given as
\begin{equation}
H/\hbar=\Delta \left(a^{\dag}a+b^{\dag}b\right)+G\left(a^{\dag }b+b^{\dag }a\right)+i\sqrt{\eta \kappa}(\varepsilon a^{\dag }-\varepsilon^{\ast}a). \label{Hamiltonian}
\end{equation}
The cavity mode is described by the bosonic \MSB{operators} $a$ ($a^{\dag}$) and the mechanical mode is described by $b$ ($b^{\dag}$), respectively. The detuning is defined as $\Delta=\omega_{c}-\Omega_{p}$. \vari{Here, $G=g\sqrt{\bar{n}_{\textrm{c}}}$ is the linearized and field-enhanced coupling strength. The quantity $g$ is the single-photon optomechanical coupling rate and $\bar{n}_{\textrm{c}}$ is the average number of photons in the cavity.} Thus, the cavity-mechanics coupling rate $G$ can be continuously adjusted by controlling the power of the coherent pump tone. Without loss of generality, $G$ is assumed to be a real number. The coupling to \MSB{the} SMA connector results in an external decay rate of $\kappa_{e}$. The cavity mode can be characterized by a total loss $\kappa$ and the external coupling parameter $\eta=\kappa_{e}/\kappa$. By considering the input-output relations, the transmission coefficient is given as
\begin{equation}
t=1-\frac{\eta \kappa \left(i\Delta +\gamma_{m}/2\right)}{\left(i\Delta +\gamma_{m}/2\right) \left( i\Delta +\kappa/2\right)+G^{2}}, \label{Transimission}
\end{equation}
where $\gamma_{m}$ represents the decay rate of the mechanical mode. The amplitude and phase responses are given as $T=|t|^2$, and $\varphi=\arg \left(t\right)$, respectively. The transmission at cavity resonance with zero-detuning (setting $\Delta$ to be zero) becomes
\begin{equation}
t_{z}=\frac{G^{2}-\left(\eta-1/2\right) \kappa \gamma_{m}/2}{G^{2}+\kappa \gamma_{m}/4}. \label{TRcentere}
\end{equation}

At zero-detuning, $t_{z}$ is always a real number. Setting $t_{z}$ to be zero yields a critical coupling $G_{\mathrm{c}}=\sqrt{\left(\eta-1/2\right) \kappa \gamma_{m}/2}$. When $G<G_{\mathrm{c}}$ ($G>G_{\mathrm{c}}$), $t_{z}$ is negative (positive). Under critical coupling, the amplitude at \MSB{the} cavity resonance becomes zero, indicating that the mechanically induced CPA (MCPA) occurs. The phase at zero detuning (cavity resonance) is given \MSB{as} $\varphi_{z}=\arg\left(t_{z}\right)$. It is notable that the imaginary part of $t_{z}$ \MSB{on resonance} is always zero [i.e., $\mathrm{Im}(t_{z})=0$]. We then have $\mathrm{tan}\left(\varphi_{z}\right)=\mathrm{Im}(t_{z})/\mathrm{Re}(t_{z})=0$, and the nontrivial solutions for the phase at zero detuning are $\varphi_{z}\in[0,\pi]$. Because $\mathrm{cos}(\varphi_{z})\varpropto\mathrm{Re}(t_{z})$, and the \MSB{value} of $\mathrm{cos}(\varphi_{z})$ is negative (positive) for $G<G_{\mathrm{c}}$ ($G>G_{\mathrm{c}}$), hence \MS{yielding} a phase $\varphi_{z}=\pi$ ($\varphi_{z}=0$), respectively. Remarkably, the MCPA mediates an abrupt $\pi$-phase transition at the critical coupling.

\textit{Observation of the MCPA and anti-lasing.}---In earlier work, CPA and anti-lasing have been achieved by controlling the interference of multiple incident waves inside a thin film~\cite{chong2010coherent,wan2011time,stone2011gobbling}. \vari{For cavity optomechanical systems, the probe light can be perfectly absorbed through the constructive interference with the down-converted pump tone from the blue-detuned sideband~\cite{hocke2012electromechanically}. \MS{In this case,} the interaction Hamiltonian is written as $H_{\textrm{int}}/\hbar=G(a^{\dag}b^{\dag}+ab)$, corresponding to parametric amplification~\cite{massel2011microwave,fong2014microwave,zhang2016cavity}. Switching from blue-detuned to red-detuned driving can \MSB{induce a change from constructive to destructive interference}, giving rise to OMIT~\cite{aspelmeyer2014cavity}. The interaction Hamiltonian becomes \MS{beam-splitter} like $H_{\textrm{int}}/\hbar=G(a^{\dag}b+b^{\dag}a)$ \MS{as in the present case, with the transmission coefficient given} in Eq.~(\ref{TRcentere}). If the \MS{cavity} is under-coupled ($\eta<1/2$), e.g., for the devices studied in Refs.~\cite{weis2010optomechanically,safavi2011electromagnetically,karuza2013optomechanically}, destructive interference occurs, and only OMIT is \MS{observed} (see detailed discussions in Sec.~IV of Ref.~\cite{Liu2021antilasing}). However, \Liu{as will be shown in this work,} \MS{with over-coupling} ($\eta>1/2$), interference becomes constructive again and OMIA can occur.}

For the experiment, the device is mounted on a cold plate with a cryogenic temperature of 10~mK inside a dilution refrigerator. The measurement setups are shown in Sec.~III of Ref.~\cite{Liu2021antilasing}. For our device, the external coupling corresponds to over-coupling, $\eta=0.651$, and the pump driving is red-detuned. \MSB{The resonance frequencies of the cavity and of the mechanical resonator} are measured to be $\omega_{c}/2\pi=5.318$~GHz and \vari{$\omega_{m}/2\pi=755.5$~kHz}, respectively. The other parameters for this device are calibrated as $\kappa/2\pi=420$~kHz, and $\gamma_{m}/2\pi=9.7$~mHz. Detailed calibration processes can be found in Sec.~V of Ref.~\cite{Liu2021antilasing}. Then, the calculated critical coupling is given as $G_{\mathrm{c}}/2\pi=17.53$~Hz.
\begin{figure}[ptb]
\includegraphics[scale=0.35]{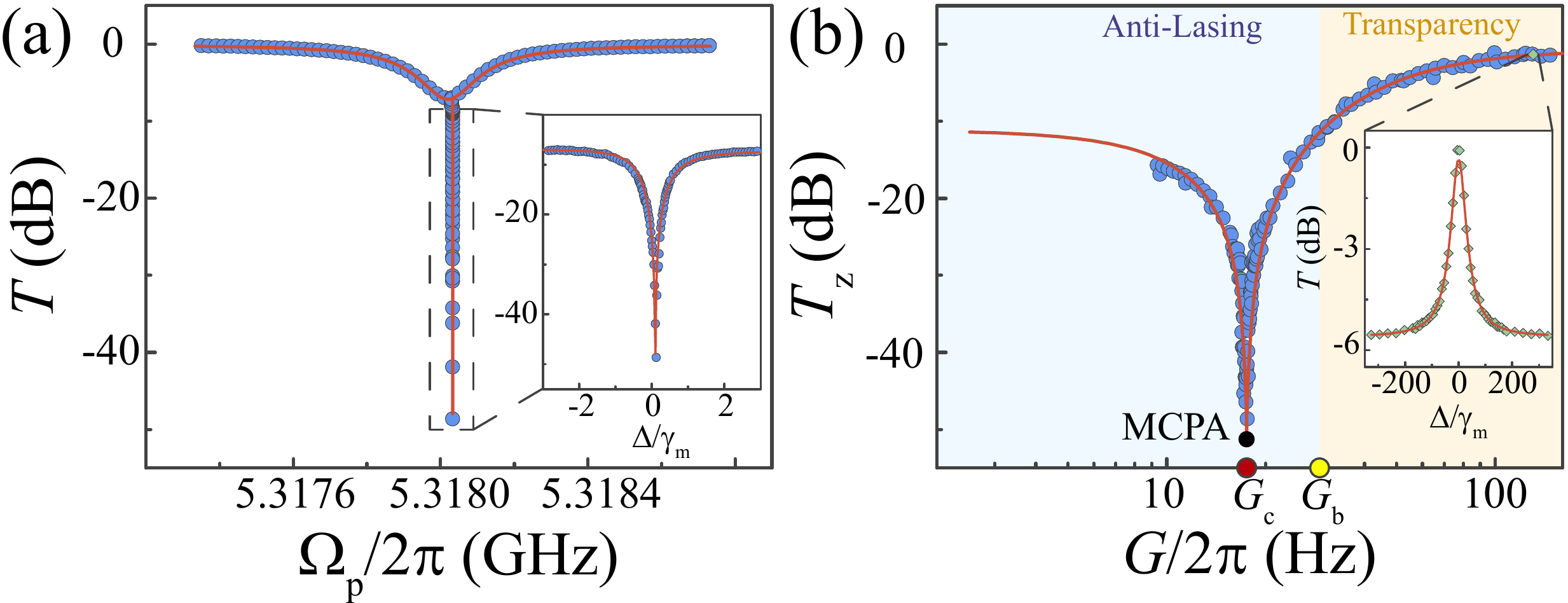}
\caption{(a) The transmission spectrum for the MCPA; and (b) The amplitude at cavity resonance versus coupling strength. Inset in (a) zooms inside the absorption dip. Inset in (b) demonstrates a typical transparent window in the transparency regime with $G/2\pi=176.8$~Hz. Blue-circle: data; Red-solid lines: fit. Blue (orange) background marks the anti-lasing (transparency) regime, respectively.}
\label{fig2}
\end{figure}

The amplitude \MSB{response}, with a coupling strength $G/2\pi=17.66$~Hz, \MSB{is} presented in Fig.~\ref{fig2}(a). The amplitude at the absorption dip is nearly 50~dB lower than the baseline of the transmission spectrum. As shown in Fig.~\ref{fig2}(b), the \MSB{amplitude} $T_{z}$ at \MSB{the} cavity resonance \MSB{is} measured with \MSB{a} continuously increasing $G$. \MSB{The} amplitude $T_{z}$ dramatically drops until $G$ reaches $G_{c}$, and then starts to increase with further increasing the coupling strength. The MCPA occurs and \Liu{anti-lasing} is observed at $G_{c}$. Figure~\ref{fig2}(b) shows that the measured critical coupling \MSB{occurs} at $G_{c}/2\pi=17.528$~Hz, which agrees well with the calculated value. The coupling strength corresponding to the boundary between \Liu{absorption (anti-lasing) and transparency} is given as $G_{b}=\sqrt{\left(\eta/2-1/4\right)\kappa\gamma_{\mathrm{m}}/\left(1-\eta\right)}$. For our device, the boundary-coupling is $G_{b}/2\pi=29.68$~Hz. The inset in Fig.~\ref{fig2}(a) [Fig.~\ref{fig2}(b)] shows the typical amplitude response in the transparency (anti-lasing) regime, respectively.

The MCPA allows for the following physical interpretation. Optomechanical interaction of the pump and probe tones creates a sideband at the probe tone frequency. Below critical coupling ($G<G_{c}$), the mechanical sideband is smaller than the probe tone, leading to only partial absorption, and the phase is that of the probe tone. At the critical coupling ($G=G_{c}$), the sideband perfectly interferes with the probe field inside the cavity, leading to coherent perfect absorption and no emitted field. Above critical coupling but below the boundary-coupling ($G_{c}<G<G_{b}$), the sideband now dominates \MSB{over} the probe, so that only part of the sideband tone is transmitted out of the cavity, \MSB{being} $\pi$ out of phase with respect to the original probe tone. Above the boundary-coupling ($G>G_{b}$), the sideband is strong enough to increase the transmission beyond the bare microwave cavity background, and a transmission peak \MSB{appears}.

\begin{figure}[ptb]
\includegraphics[scale=0.35]{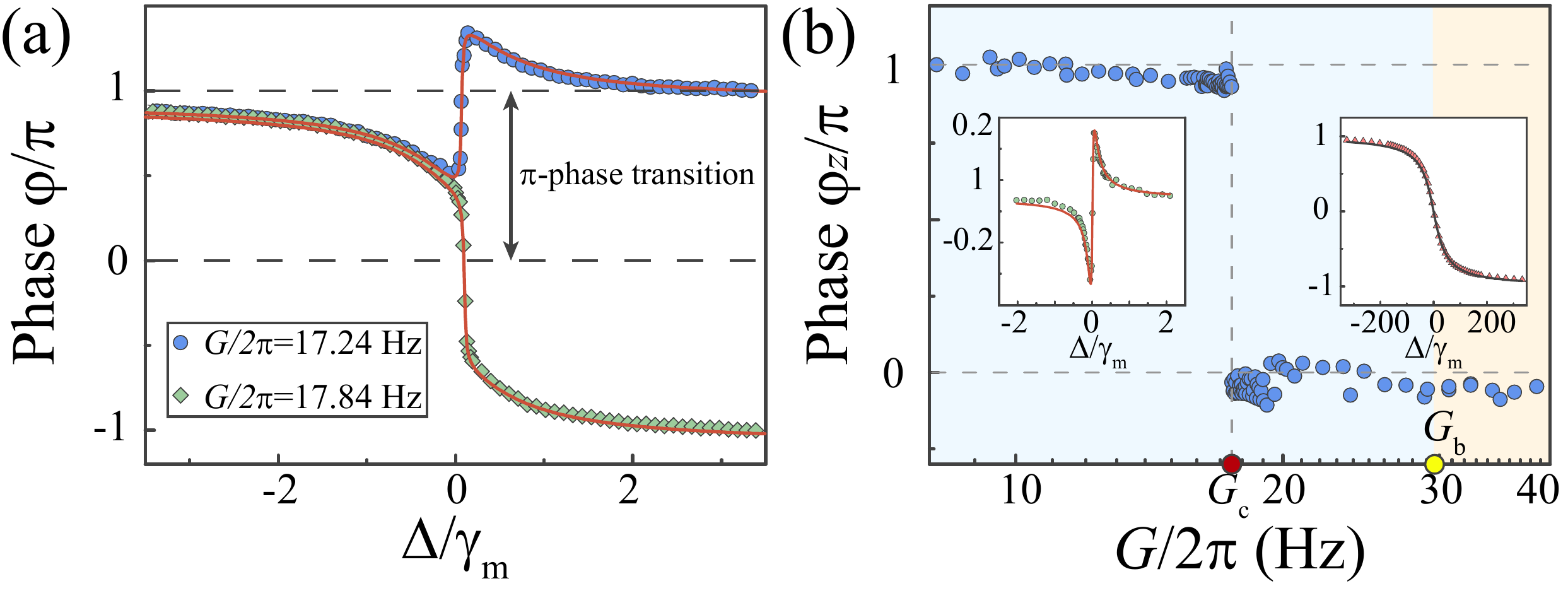}
\caption{(a) Phase $\varphi$ versus detuning $\Delta$ for two coupling strengths
close to the critical coupling $G_{c}$ on both sides. (b) Phase $\varphi_{z}$ at cavity resonance versus coupling strength $G$. Insets in (b) illustrate typical phase response with coupling strengths on either side of $G_{c}$. Red- and black-solid lines: fit; Circle, rhombus, and triangle: data.}
\label{fig3}
\end{figure}

\textit{The $\pi$-phase transition and phase slope flip.}---In the anti-lasing regime, for a specific value of $T_{z}$, there exist two coupling strengths which \MSB{are located to the left or right of} $G_{c}$, respectively. The phase responses for $G/2\pi=17.24$ and $17.84$~Hz (corresponding to $T_{z}=-42$~dB) are shown in Fig.~\ref{fig3}(a). Under these two coupling strengths, although both the amplitude responses exhibit absorption dips [as shown in Fig.~\ref{fig2}(a)], the phase at zero-detuning undergoes an abrupt shift by $\pi$ (referred to as $\pi$-phase transition). \vari{As the detuning becomes large, \MS{at} the coupling \MS{values just} above and below $G_{c}$, \MS{the phase becomes $\pi$ at over-coupling, whereas under-coupling yields a zero phase \cite{Liu2021antilasing}.} Thus, switching from \MS{under- to over-coupling leads to \MSB{the} destructive interference (OMIT) changing into constructive interference (OMIA).}}

The \MSB{phases $\varphi_{z}$} at the cavity resonance are measured for different coupling strengths $G$ and are shown in Fig.~\ref{fig3}(b). An obvious $\pi$-phase transition occurs at the critical \MSB{coupling $G_{c}$, where} all the inject microwave photons are absorbed ($T_{z}$ becomes zero). Thus, the MCPA mediates such $\pi$-phase transition. \MSB{The} insets in Fig.~\ref{fig3}(b) show a typical phase response, where $G$ is located either on the left or right side of $G_{c}$. On the right side of $G_{c}$, it is notable that the phase response maintains the same lineshape [as shown in the right inset in Fig.~\ref{fig3}(b)] in both the anti-lasing and transparency regimes. It is \MS{also worthy to note} that the $\pi$-phase transition further flips the slope in the frequency dependence of \MSB{the} phase, as shown in Fig.~\ref{fig3}(a). For $G<G_{c}$ ($G>G_{c}$), the \vari{phase slope} at zero-detuning is always positive (negative), respectively, undergoing a positive-negative transition around the cavity resonance. Thus, the phase slope at $G_{c}$ is discontinuous, which indicates a singularity for the phase slope.

\textit{Infinite discontinuity and a singularity in the group delay.}---Group delay is given by $\tau = \partial \varphi /\partial \Omega_{p} = - \partial \varphi /\partial \Delta$. Hence, multiplying the phase slope by a negative-sign implies a group delay. Thus, a singularity exists in the group velocity \MS{in our case}. In the experiment, the group delay is initially measured at several frequencies of the probe-tone, which is generated and further analyzed by \MSB{a} VNA.

\begin{figure}[ptb]
\includegraphics[scale=0.35]{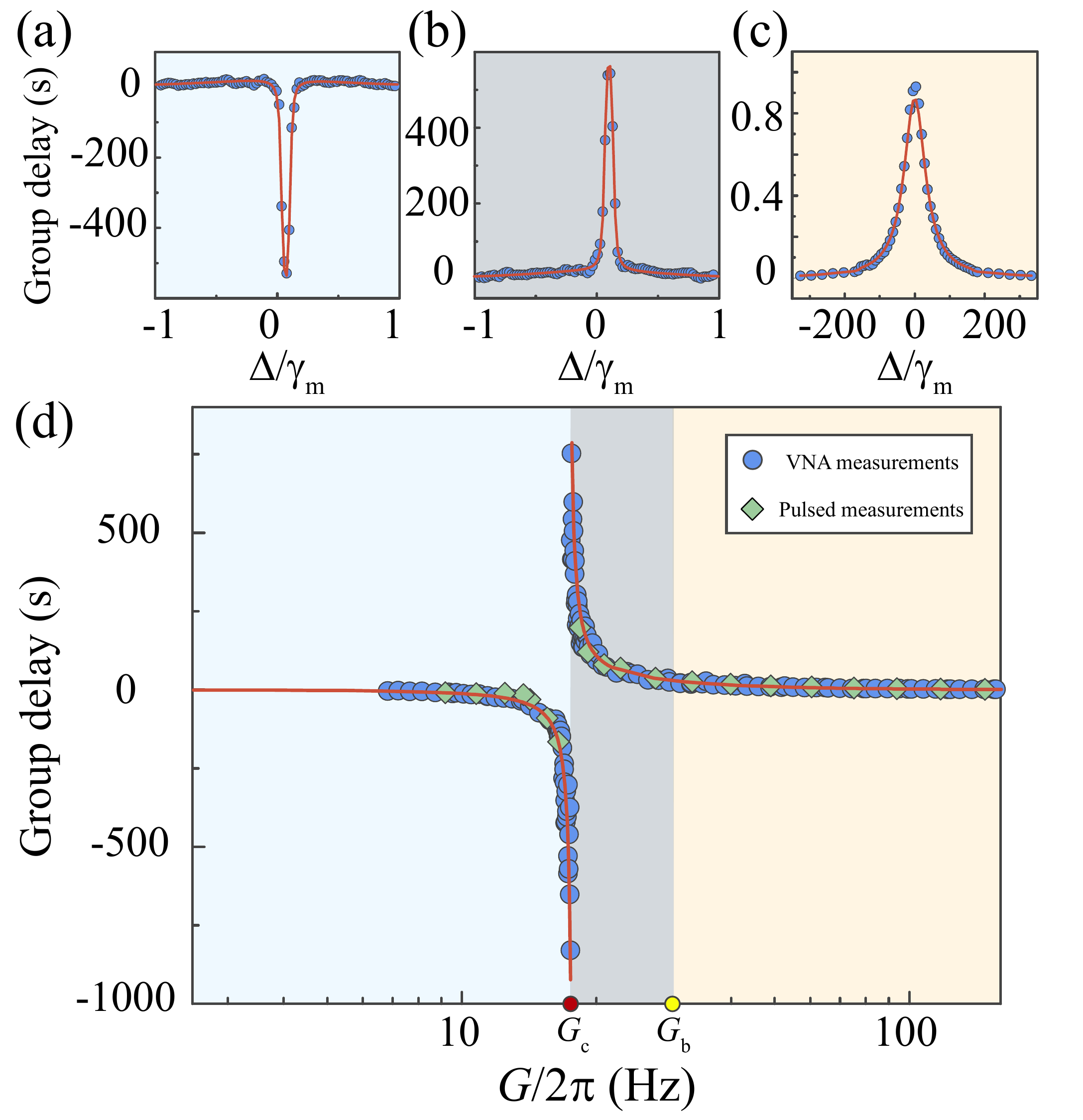}
\caption{\MS{Group} delay $\tau$ versus detuning $\Delta$ are shown in (a) with $G/2\pi$=17.24~Hz, (b) with $G/2\pi$=17.84~Hz, and (c) with $G/2\pi$=176.8~Hz, respectively. (d) The group delay at zero-detuning ($\tau_{z}$) versus coupling strength $G$ are measured. Blue-circle (green-rhombus) marks the VNA (pulsed) measurements. Red-solid: fit.}
\label{fig4}
\end{figure}

Figure~\ref{fig4}(a), (b) and (c) \MSB{shows} the measured group delay $\tau$ for three different coupling strengths. The corresponding phase responses have been presented in Fig.~\ref{fig3}(a), and right inset in Fig.~\ref{fig3}(b), respectively. Figure~\ref{fig4}(a) shows group advance with a coupling $G/2\pi=17.24$~Hz, and Fig.~\ref{fig4}(b) shows group delay with a coupling $G/2\pi=17.84$~Hz. These two coupling strengths \MSB{are} inside the anti-lasing regime \MSB{and} very close to $G_{c}$ from either side. However, the amplitude responses around $G_{c}$ only exhibit absorption dips, as shown in Fig.~\ref{fig2}(a). Figure \ref{fig4}(c) shows group delay with a coupling $G>G_{b}$ in the transparency regime, and its corresponding amplitude response has been shown in the inset of Fig.~\ref{fig2}(b).

The group delay at \MSB{the} cavity resonance $\tau_{z}$ versus different coupling strengths \MSB{is} demonstrated in Fig.~\ref{fig4}(d). Fast (slow) light with negative (positive) group delay are observed on the left (right) sides of $G_{c}$, respectively. The group advance and delay approach infinite at $G_{c}$. Thus, the $\pi$-phase transition and \MSB{the} phase slope flip result in a singularity in group delay with an infinite discontinuity, \MSB{located} at $G_{c}$. Compared to group delay in the transparency (or OMIT) regime extensively discussed in the literature~\cite{weis2010optomechanically,safavi2011electromagnetically,karuza2013optomechanically,dong2013transient,dong2015brillouin,kim2015non,teufel2011circuit,ockeloen2019sideband,massel2012multimode,zhou2013slowing,fan2015cascaded}, the group delay around the singularity is greatly enhanced or even diverges.

\textit{Pulsed measurements.}---To directly explore the group delay, microwave pulses are generated by modulating the amplitude of \MSB{a} probe tone derived from a microwave generator. The Gaussian-shaped envelopes are generated with an arbitrary waveform generator. To avoid any pulse distortion in group delay measurements, the pulse bandwidths are chosen to be narrower than the transparency (absorption) window width. The emission of a probe pulse is synchronized with the acquisition of the transmitted probe field \cite{Liu2021antilasing}.
The group delay is given by the difference between measured pulse center-time (time at peak-center) with and without a continuous-wave pump tone sent into the cavity.
\begin{figure}[ptb]
\includegraphics[scale=0.3]{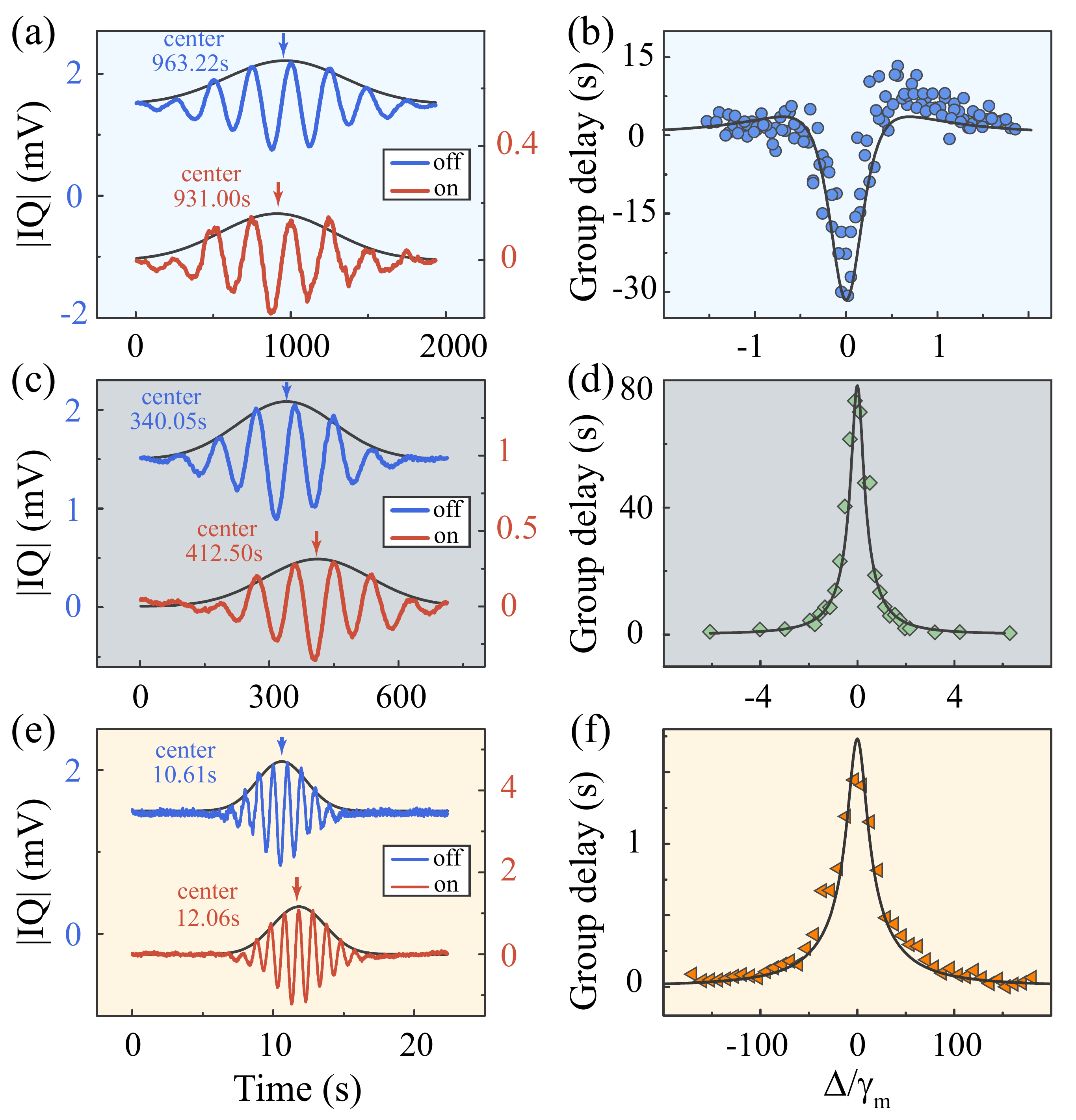}
\caption{Pulses propagating at cavity resonance are shown in (a), (c) and (e) with different coupling strengths. The extracted group delay versus $\Delta$ are given in (b), (d), and (f), respectively. The coupling strength is fixed at $G/2\pi$=11.87~Hz for (a) and (b); $G/2\pi$=23.93~Hz for (c) and (d);  $G/2\pi$=155.1~Hz for (e) and (f). Red-solid (blue-solid) Gaussian-shaped pulse is measured with (without) the pump tone. Black-solid: fit.}
\label{fig5}
\end{figure}

In the anti-lasing regime, a group advance [as shown in Fig.~\ref{fig5}(a) and (b)] is achieved, when $G$ is located on the left side of the singularity (e.g., $G/2\pi=11.87$~Hz). Figure~\ref{fig5}(c) and (d) \MSB{shows} a typical group delay when $G$ is to the right from the singularity (e.g., $G/2\pi=23.93$~Hz). In the transparency regime, a regular group delay is measured [as shown in figure~\ref{fig5}(c) and (d)] with, e.g, $G/2\pi=155.1$~Hz. The maximum group delay ($\tau_{m}$) occurs at the zero detuning point.

It is notable that the maximum group delay is proportional to the mechanical lifetime (i.e., $\tau_{m}\propto1/\gamma_{m}$) in \MSB{the} transparency regime~\cite{safavi2011electromagnetically,zhou2013slowing,fan2015cascaded}. Assisted by \MSB{a} low mechanical damping ($\gamma_{m}=9.7$~mHz), the measured maximum group delay in Fig.~\ref{fig5}(f) arrives at $\tau_{m}=1.2$~s, which is \MSB{an improvement of} eight orders of magnitude \MSB{compared to} 50~ns \MSB{measured} in optomechanical crystals, e.g., in Ref.~\cite{safavi2011electromagnetically} and three orders of magnitude improvement over 3~ms measured earlier in circuit electromechanics~\cite{zhou2013slowing}. The measured group advance and delay when probe-tone detuning is varied, \MSB{exhibit} an excellent agreement with our model. The measured group delay in a pulsed measurement also agrees well with the VNA group delay measurements (see Sec.~VI of Ref.~\cite{Liu2021antilasing}).

\textit{Conclusion.}---\MSB{We experimentally study anti-lasing and MCPA, and find the latter occurs at a critical optomechanical coupling.} Meanwhile, MCPA (anti-lasing) mediates a $\pi$-phase transition, which further flips the phase slope at cavity resonance. Subsequently, the group delay measured with \MSB{a} continuous or pulsed probe field exhibits a singularity, around which the group advance and delay diverge. An abrupt transition from \MSB{an} infinite group advance to delay is observed at this singularity. \vari{The $\pi$-phase transition and group-delay singularity could be further explored within an amplification regime by, e.g., utilizing blue-detuned sideband driving, to overcome the absorption loss~\cite{Liu2021antilasing}.} The singularity with \MSB{the} infinite group delay may motivate ways to realize extremely \MSB{long signal storage times as an alternative} to the schemes based on, e.g., solid nuclear spins~\cite{zhong2015optically}, or atomic frequency comb~\cite{ma2021one}. Our scheme could also be realized in several other cavity-based quantum systems such as the cavity spintronics~\cite{wang2020dissipative}, circuit-QED~\cite{haroche2020cavity}, and microcavity-photonics~\cite{chen2018nonreciprocity}.

\textit{Acknowledgements.}---This work is supported by the National Key Research and Development Program of China (Grant No. 2016YFA0301200), the National Natural Science Foundation of China (Grant No. 12004044, Grants No. 62074091, and No. U1930402), and Science Challenge Project (Grant No. TZ2018003), and by the Academy of Finland (contracts 307757, 312057, 336810).

%

\end{document}